\documentclass[a4paper,fleqn,usenatbib]{mnras}

\usepackage{newtxtext,newtxmath}
\usepackage[T1]{fontenc}
\usepackage{ae,aecompl}

\usepackage{graphicx}	% Including figure files
\usepackage{amsmath}	% Advanced maths commands
\usepackage{amssymb}	% Extra maths symbols
\usepackage{multirow}
\usepackage{lscape}

\title[Optical and X-ray Observations of NGC 1569]{Optical and X-ray Observational Search for the Possible Supernova Remnant Candidates in the Nearby Galaxy NGC 1569}

\author[Ercan et al.]{
E. N. Ercan,$^{1}$\thanks{E-mail: ercan@boun.edu.tr (NE)}
E. Aktekin,$^{1}$
N. Cesur$^{2}$
and A. T\"{u}mer$^{1}$
\\
$^{1}$Department of Physics, Bogazici University, 34342, Istanbul, Turkey\\
$^{2}$Department of Physics, Y{\i}ld{\i}z Technical University, 34220, Istanbul, Turkey\\
}

\date{Accepted XXX. Received YYY; in original form ZZZ}

\pubyear{2018}

\begin{document}
\label{firstpage}
\pagerange{\pageref{firstpage}--\pageref{lastpage}}
\maketitle

\begin{abstract}
We report the results of our investigation on the possible existence of supernova remnants (SNRs) in the nearby galaxy, NGC 1569, using the CCD imaging and spectroscopic observations from the RTT150 telescope of TUG-TUBITAK in Antalya, Turkey for two different observing periods. Using [S\,{\sc ii}]/H$\alpha$ $\geq$ 0.4 standard criteria, the identification of 13 new SNR candidates for this galaxy is proposed for two different epochs. We found the [S\,{\sc ii}]/H$\alpha$ ratios ranging from 0.46--0.84 and H$\alpha$ intensities ranging from (2.2--32) $\times$ 10$^{-15}$ erg cm$^{-2}$ s$^{-1}$. [S\,{\sc ii}]$\lambda\lambda$6716/6731 average flux ratio is calculated from the optical spectra for only one possible SNR candidate. By using this ratio the electron density, $N_{\rm e}$, is estimated to be 121 $\pm$ 17 cm$^{-3}$ and by using the [O\,{\sc iii}]$\lambda$5007/H$\beta$ ratio of the same spectrum, the shock wave velocity, $V_{\rm s}$, is estimated to be between 100 < Vs < 150 km s$^{-1}$. Using {\it Chandra} data, we find that out of 13 SNR candidates only 10 of them have yielded a spectrum with good statistics, confirming the existence of  10 SNR candidates with matched positions in X-ray region as well. We measure the 0.5--2 keV band flux down to 0.58 $\times$ 10$^{-15}$ erg cm$^{-2}$ s$^{-1}$ for 10 X-ray sources. Our spectral analysis revealed that spectra of the SNR candidates are best modelled with the Collisional Ionisation Equilibrium plasma with a temperature range of 0.84 keV < \textit{kT$_{e}$} < 1.36 keV.
\end{abstract}

\begin{keywords}
galaxies: individual NGC 1569, galaxies: star formation - ISM: supernova remnants - X-rays: galaxies - X-rays: ISM

\end{keywords}

%%%%%%%%%%%%%%%%%%%%%%%%%%%%%%%%%%%%%%%%%%%%%%%%%%

%%%%%%%%%%%%%%%%% BODY OF PAPER %%%%%%%%%%%%%%%%%%

\section{INTRODUCTION}

The optical, X-ray and radio investigations of extragalactic supernova remnants (SNRs) can be used to understand the evolution of SNRs, to estimate the supernovae rate and SNRs population in galaxies, and to study the global properties of the  interstellar  medium and environments in the galaxy \citep[see e.g.][]{matonickandfesen1997, pannutietal2000, sonbasetal2010, leeandlee2014}.

Strong ratio of [S\,{\sc ii}] with respect to H$\alpha$ line of SNRs are great indicators of this nebula emission which also has an extended X-ray along with powerful non-thermal radio emission \citep{long1985}. In order to identify SNRs,  investigation has to be performed by the utilization of tools such as narrow band imaging with H$\alpha$ and [S\,{\sc ii}] filters, radio observations, and X-ray imaging.

NGC 1569 is a nearby dwarf irregular and a well-known starburst galaxy found in the Camelopardalis constellation, which was discovered by William Herschel in 1788. With an apparent visual magnitude of 11.9, NGC 1569 lies at a distance of 11 million light years, i.e. 3.4 Mpc \citep{lellietal2014, jacksonetal2011, kobulnickyandskillman1997}.

\citet{grocholskietal2008} have presented deep Hubble Space Telescope (HST) Advanced Camera for Surveys / Wide Field Channel (ACS/WFC) results and determined the distance to be 3.36 $\pm$ 0.20 Mpc which is considerably farther away than its previously observed value of only 2.2 $\pm$ 0.6 Mpc by \citet{israel1988}. It is part of the group as IC 342 group of galaxies, which was initially assumed to be an isolated galaxy due to its underestimated distance. Its high density character reveals NGC 1569's starburst nature, which in general is the result of galaxy interactions. In addition, as opposed to what was originally thought, this galaxy contains higher numbers of starbursts implied by their extended spread. Early predictions for star formation rates for stars younger than 1 Gyr enhanced the evidence by a factor of more than two, while older stars were not subjected to such restrictions.

Another work done in optical for the intrinsic extinction distribution of NGC 1569, \citet{relanoetal2006a}, reported the use of an optical extinction map ($A_{\rm v}$) obtained from their H$\alpha$/H$\beta$ emission line ratio for this purpose, and they obtained values up to $A_{\rm v}$ = 0.8 mag.

\citet{devostetal1997} reported deep H$\alpha$ imaging and multi-slit optical spectroscopy results of 16 H\,{\sc ii} regions. The optical extinction for NGC 1569 was obtained from the Balmer H$\alpha$/H$\beta$ line ratio emphasizing the fact that most of the observed $A_{\rm v}$ was caused by Milky Way emission, which gave ($A_{\rm v}$) local = 1.61 $\pm$ 0.09. However, the intrinsic $A_{\rm v}$ for NGC 1569 was found to be 0.65 $\pm$ 0.04.

\citet{chomiukandwilcots2009} reported their survey for 4 nearby irregular galaxies for SNR candidates in the radio region including NGC 1569, by using deep (1$\sigma \sim 20 \mu$Jy) high resolution ($\sim$20 pc) VLA continuum data at 20, 6, and 3.6 cm. They identified discrete sources and categorized them as SNRs, H\,{\sc ii} regions, or background radio galaxies with the use of radio spectral indices and H$\alpha$ images. Their classifications were found to be consistent with the early studies. According to their results NGC 1569 contains 23 SNR candidates.

\citet{dellacecaetal1996} have studied the X-ray aspects of dwarf starburst inside NGC 1569, using their {\it ASCA} observations along with archival data from {\it ROSAT} PSPC and HRI, H$\alpha$ images of optical broad and narrow bands, combined with recent IR K-band images. Soft and hard components were discovered from the {\it ASCA} SIS broad-band X-ray spectrum (0.5--6 keV). Thermal model fitting of the soft component provided the temperature values ranging from $\sim$0.64--0.8 keV, whereas $\sim$3.7 keV thermal model or 2.1 photon index of power-law component individually fitted the hard component. Extended X-ray emission was observed in the soft component and its morphology was associated with H$\alpha$ filaments.

It is generally believed that X-ray observations may provide strong evidence that starbursts can drive gas out of the dwarf galaxies, as it was seen from the diffuse X-ray emission analysis of NGC 1569. The X-ray data can establish the diffuse X-ray emission in NGC 1569 produced by hot gas and also can tell us that the temperature of this gas may exceed the depth of the galaxy's shallow potential well.

\citet{jurgenetal2005} also found an extended diffuse X-ray emission which was associated with temperatures ranged from 1.6 to 7.6 $\times 10^{6}$ K from {\it Chandra} X-ray observations.

The most recent optical and X-ray observations of NGC 1569 have been reported by \citet{sánchez-crucesetal2015} describing the optical lines of H$\alpha$ and [S\,{\sc ii}]$\lambda\lambda$6716,6731 \AA. They achieved their optical observations in H$\alpha$ and [S\,{\sc ii}] conducted with the UNAM scanning FabryPerot interferometer (PUMA) and the X-ray data obtained from the {\it Chandra} data archive. They observed several superbubbles, filaments, and supershells in NGC 1569 for which they determined their sizes as well as their kinematical properties. In their {\it Chandra} observation result also a total of 54 possible X-ray emission coordinates  were recorded  in their Table 9. Amongst 54 of  them  they listed  a possible description of their characteristic types that could be considered as  active galactic nuclei, X-ray sources, X-ray binaries, stars and two possible SNRs.

The main physical properties of NGC 1569 retrieved from aforementioned studies are summarised in Table~\ref{tab:properties}.

\begin{table}
\centering
  \caption{Physical properties of NGC 1569.}
  \label{tab:properties}
  \begin{tabular}{@{}ccc@{}}
\hline
      Parameters & Value & References \\
      \hline
     Other Names& UGC03056, ARP210, VII Zw 016 & (1)\\
 Morphological Type & IBm  & (1)\\
\multirow{2}{*}{Position (J2000)}&$\rmn{RA}(J2000)$ : $04^{\rmn{h}} 30^{\rmn{m}} 47^{\rmn{s}}$.48  & \\
 &$\rmn{Dec.}~(J2000)$: $64\degr 50\arcmin 55\arcsec$.70 & \\
 Visual Magnitude& 11.9& (2)\\
 Distance& 3.4 Mpc & (2) \\
   Diameter& 1.69 kpc& (3)\\
   \hline
  \end{tabular}
 References: (1) NASA Extragalactic Database, (2) \citep{lellietal2014, jacksonetal2011, kobulnickyandskillman1997}, (3) \citet{sánchez-crucesetal2015}.
\end{table}

We have particular interest in performing a systematic study to search for possible SNR candidates in nearby galaxies and we started with NGC 1569 for which we had long exposure times in optical in two different epochs as well as the X-ray archival data of {\it Chandra} (see Section 2 for both) to be able to achieve our aim. This work will present the first optical study for NGC 1569 using the data of RTT150 at TUG.

Some of the possible SNR candidates have already been observed in optical region by us earlier (NGC 1569 was one of them) but the limited observing times did not let us present our results with poor time statistics then. However, even with short time observations we have noticed and detected some SNR candidates in our preliminary results (including so-called No.1 observations, in this text) but preferred to wait for another chance of getting a better and longer observing time to observe NGC 1569 again with RTT150.

In fact, we finally achieved this with No.1 and a longer No.2 observations. As we have mentioned in the text during the No.1. (spectral) observations, we had only one chance of getting the spectrum of one of the SNR candidate out of a total of 13 SNR candidates for NGC 1569 because of the limited observing conditions. 
Having much longer observing period during the second epoch of our observations, we were then able to present both of our observation in a total "combined" observational optical data obtained from the addition of No.1 and and No.2 as shown in Table 2.

\section{OBSERVATIONS AND DATA REDUCTION}

Our optical observations are performed in two different periods of observing time of RTT150\footnote{Specifications of RTT150 and TFOSC are available at \href{http://www.tug.tubitak.gov.tr}{http://www.tug.tubitak.gov.tr}.} approximately seven months apart from each other. So called-No.1 is done on the 28 $\&$ 29th March 2017 and No.2 is on the 25th December 2017.

We have combined our so-called No.1 and No.2 observations as one single long exposure optical observation by RTT150. One can see the CCD configurations of our observations through the link provided at the TUG Observatory internet page : \href{http://tug.tubitak.gov.tr/en/teleskoplar/rtt150-0}{http://tug.tubitak.gov.tr/en/teleskoplar/rtt150-0}.

\subsection{Optical Imaging Observations}

The optical imaging observations of NGC 1569 took place on the 29th March 2017 (No.1 observation) and 25th December 2017 (No.2 observation) with the 1.5 m RTT150 in Antalya, Turkey and images are obtained with the low-resolution TUG Faint Object Spectrograph and Camera (TFOSC). It is equipped with a 2048 $\times$ 2048 back-illuminated camera. For the No.1 observations, a pixel size of 15 $\mu$m corresponding to a 13.5 arcmin square field of view (FoV), and for the No.2 observations, a pixel size of 13.5 $\mu$m was used giving a 11.5 arcmin square FoV attached to the Cassegrain focus of the RTT150. Narrow-band filters centered lines of [S\,{\sc ii}], H$\alpha$, and continuum filters were used. The interference filter characteristics and our exposure times of imaging observations are shown in Table~\ref{tab:observation}.

\begin{table*}
\centering
  \caption{Combined (No.1 and No.2) optical observations: The interference filter characteristics and our exposure times of imaging observations.}
  \label{tab:observation}
  \begin{tabular}{@{}ccccc@{}}
  \hline
 Filter & Wavelength  & FWHM &Exposure times & Obs. date \\
  & (\AA) & (\AA)& (s)&\\
  \hline
  H$\alpha$ & 6563 & 80 & 3 $\times$ 300&2017/03/28-29\\
   &  &  & 6 $\times$ 900 & 2017/12/25\\
 H$\alpha$ cont. & 6446 & 123 & 1 $\times$ 300&2017/03/28-29\\
 &  &  & 6 $\times$ 300 & 2017/12/25\\
 $[$S$\,${\sc ii}$]$ & 6728 & 54& 3 $\times$ 300&2017/03/28-29\\
 &  &  & 6 $\times$ 900 & 2017/12/25\\
 $[$S$\,${\sc ii}$]$ cont. & 6964 & 350 & 1 $\times$ 300&2017/03/28-29\\
  &  &  & 6 $\times$ 300 & 2017/12/25\\
\hline
\end{tabular}
\end{table*}

\begin{figure}
\centering
  \vspace*{17pt}
 \includegraphics[width=18cm]{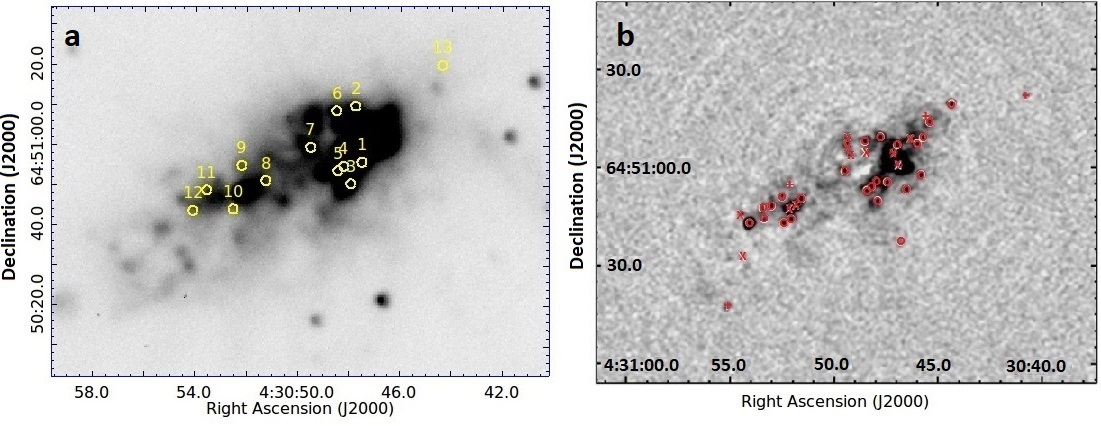}
  \caption{(a): Optical Combined  Data:The H$\alpha$ - H$\alpha$ continuum image of NGC 1569 with J2000 equatorial coordinates from optical Observations. SNR candidates are marked with yellow circles and with their ID numbers (in total 13) given in Table~\ref{tab:optic_result}. (b)from \citet{chomiukandwilcots2009} 21 cm radio map of the central region of NGC 1569, labeled with J2000 equatorial coordinates. Here, + signs indicate background galaxies, circles are their SNR candidates, x signs are their  H\,{\sc ii} regions, and * signs are sources which may be H\,{\sc ii} regions or SNR candidates. Seven sources fall outside the bounds presumed to be all classified as background galaxies.}
  \label{fig:optic_image}
\end{figure}

Eventually Image Reduction and Analysis Facility (IRAF) was used to reduce the combined optical data. Data reduction was achieved by using standard procedures, including the corrections for bias, overscan at field, and removal of cosmic rays using the IRAF CCDPROC package. H$\alpha$ and [S\,{\sc ii}] images were subtracted from H$\alpha$ and [S\,{\sc ii}] continuum. Using {\it imarith} package of IRAF, continuum substraction and image division were realised. {\it Imarith} package which subtracts and divides the number of photons in each pixel of the obtained image. Each night the spectrophotometric standard star Feige 34 from \citet{masseyetal1988} was observed for the flux calibration. \citet{jacobyetal1987} described the conversion from instrumental counts in to physical units (erg cm$^{-2}$ s$^{-1}$). We used {\it Combine} package of IRAF to combine No.1 and No.2 observations. ([S\,{\sc ii}] - [S\,{\sc ii}] continuum) / (H$\alpha$ - H$\alpha$ continuum) combined images were obtained, then compared with the theoretical models which generally predict [S\,{\sc ii}]/H$\alpha$ ratios to be in the range of 0.5--1.0 for SNR candidates \citep[see e.g.][]{raymond1979, shullandmckee1979}. A total of 13 SNR candidates near the central part of the NGC 1569 with appropriate [S\,{\sc ii}]/H$\alpha$ values are reported as shown in Table~\ref{tab:optic_result} (see in Fig.~\ref{fig:optic_image}).

\begin{table*}
\centering
  \caption{Combined optical observations with a total of 13 SNR candidates near the central part of NGC 1569 with H$\alpha$ flux sensitivity;
appropriate [S\,{\sc ii}]/H$\alpha$ ratios and their errors in parenthesis. IDs(*)indicates the possible SNR candidates coincident with those reported in
radio by \citet{chomiukandwilcots2009} and IDs(+) indicate the possible SNR candidates, coincident with those reported in radio by \citet{greveetal2002}.}
  \label{tab:optic_result}
  \begin{tabular}{@{}ccccc@{}}
\hline
      ID & $\rmn{RA}(\rm J2000)$  & $\rmn{Dec.}~(\rm J2000)$  & [S$\,${\sc ii}$]$ / H$\alpha$ & I (H$\alpha) \times 10^{-15}$\\
      & ({$\rmn{h}$} {$\rmn{m}$} {$\rmn{s}$} )& ($\degr$ $\arcmin$ $\arcsec$) && (erg cm$^{-2}$ s$^{-1}$)\\
 \hline
 1$^*$ & $04^{\rmn{h}} 30^{\rmn{m}} 47^{\rmn{s}}$.48  & $64\degr 50\arcmin 55\arcsec$.70 & 0.82 ($\pm 0.1$) & 6.4\\
 2$^*$ & $04^{\rmn{h}} 30^{\rmn{m}} 47^{\rmn{s}}$.73  & $64\degr 51\arcmin 09\arcsec$.70 & 0.56 ($\pm 0.1$) & 7.2\\
 3$^*$ & $04^{\rmn{h}} 30^{\rmn{m}} 47^{\rmn{s}}$.91  & $64\degr 50\arcmin 50\arcsec$.30 & 0.64 ($\pm 0.2$) & 3.7\\
 4$^*$ & $04^{\rmn{h}} 30^{\rmn{m}} 48^{\rmn{s}}$.20  & $64\degr 50\arcmin 54\arcsec$.70 & 0.58 ($\pm 0.1$) & 10\\
 5$^*$ & $04^{\rmn{h}} 30^{\rmn{m}} 48^{\rmn{s}}$.42  & $64\degr 50\arcmin 53\arcsec$.60 & 0.46 ($\pm 0.2$) & 24\\
 6$^*$ & $04^{\rmn{h}} 30^{\rmn{m}} 48^{\rmn{s}}$.48  & $64\degr 51\arcmin 08\arcsec$.50 & 0.62 ($\pm 0.1$) & 6.5\\
 7$^{*+}$ & $04^{\rmn{h}} 30^{\rmn{m}} 49^{\rmn{s}}$.50  & $64\degr 50\arcmin 59\arcsec$.40 & 0.56 ($\pm 0.1$) & 7.1\\
 8 & $04^{\rmn{h}} 30^{\rmn{m}} 51^{\rmn{s}}$.24  & $64\degr 50\arcmin 51\arcsec$.05  & 0.50 ($\pm 0.1$) & 6.1\\
 9$^*$ & $04^{\rmn{h}} 30^{\rmn{m}} 52^{\rmn{s}}$.19  & $64\degr 50\arcmin 54\arcsec$.80 & 0.68 ($\pm 0.1$) & 31\\
 10 & $04^{\rmn{h}} 30^{\rmn{m}} 52^{\rmn{s}}$.51  & $64\degr 50\arcmin 43\arcsec$.86 & 0.53 ($\pm 0.1$) & 2.4\\
 11 & $04^{\rmn{h}} 30^{\rmn{m}} 53^{\rmn{s}}$.55  & $64\degr 50\arcmin 48\arcsec$.68 & 0.78 ($\pm 0.1$) & 2.3\\
 12$^*$ & $04^{\rmn{h}} 30^{\rmn{m}} 54^{\rmn{s}}$.08  & $64\degr 50\arcmin 43\arcsec$.50  &0.57 ($\pm 0.2$)&7.5\\
13$^*$ & $04^{\rmn{h}} 30^{\rmn{m}} 44^{\rmn{s}}$.35  & $64\degr 51\arcmin 20\arcsec$.01  & 0.51 ($\pm 0.2$)&6.3\\
 \hline
\end{tabular}
\end{table*}

\subsection{Optical Spectral Analysis}

The optical spectral observations were performed on the 28th of March 2017 with RTT150. High-resolution long-slit spectra of one out of 13 SNR candidates located at $\rmn{RA}(\rm J2000)=04^{\rmn{h}} 30^{\rmn{m}} 48^{\rmn{s}}$.42, $\rmn{Dec.}~(\rm J2000)=64\degr 50\arcmin 53\arcsec$.60 were obtained with the low-resolution TFOSC. Spectral analysis for only one of the SNR candidates is reported here, because of the limited seeing conditions of the telescope for the rest of the candidates.

\begin{figure}
\centering
  \vspace*{17pt}
 \includegraphics[width=15cm]{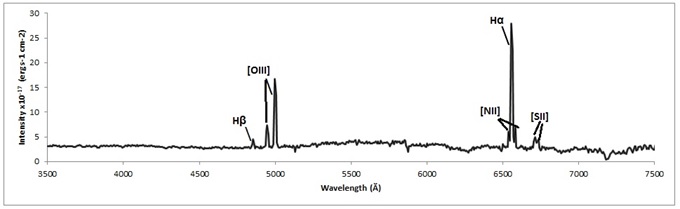}
  \caption{Optical spectrum of SNR candidate ID5; at a range of $\lambda$3500-7500 {\AA} at $\rmn{RA}(\rm J2000)=04^{\rmn{h}} 30^{\rmn{m}} 48^{\rmn{s}}$.42, $\rmn{Dec.}~(\rm J2000)=64\degr 50\arcmin 53\arcsec$.60. The Balmer H$\alpha$ $\lambda$6563{\AA}, H$\beta$ $\lambda$4861{\AA} and forbidden lines [OIII] $\lambda\lambda$4959,5007{\AA}, [NII] $\lambda\lambda$6548,6584{\AA}, [SII] $\lambda\lambda$6717,6731{\AA} can be seen in the spectrum.}
  \label{fig:optical_spectrum}
\end{figure}

During the observations, grism 15 with a dispersion of 8 {\AA} ($\lambda$3230--9120 {\AA}) was used. The slit width was 134 micron. Fe-Ar calibration lamp frames were obtained for slit width for each observation. The exposure time was 900 sec for the frame. The data reduction (the data and lamp were bias-corrected with a median bias frame, and divided by a normalised field), wavelength and flux calibrations were carried out by using the standard IRAF routines. The signal to noise ratio was calculated to be 15 at [O\,{\sc iii}]$\lambda$ 5007 {\AA} line. The signal to noise ratio was found to be low (as low as $\sim$2; longward of 7500 {\AA}) since  grism 15 with a dispersion of 8 {\AA} (3230--9120 {\AA}) was used  during our observations; so we had to cut the spectrum after 7500 {\AA}. Optical spectrum of SNR candidate ID5 ($\rmn{RA}(\rm J2000)=04^{\rmn{h}} 30^{\rmn{m}} 48^{\rmn{s}}$.42, $\rmn{Dec.}~(\rm J2000)=64\degr 50\arcmin 53\arcsec$.60) at a range of $\lambda$ 3500--7500 {\AA} is shown in Fig.~\ref{fig:optical_spectrum}.

[S\,{\sc ii}]/H$\alpha$ was calculated and the electron density was obtained through [S\,{\sc ii}]($\lambda$6716/$\lambda$6731) flux ratio as described by \citep{osterbrockandferland2006}. Regarding using [O\,{\sc iii}]$\lambda$5007/H$\beta$ line ratio diagnostics  to estimate the shock velocities, $V_{\rm s}$, associated with extragalactic  SNRs, one can follow  the standard works done by \citet{matonickandfesen1997} and the earlier discussions and calculations  of [O\,{\sc iii}]$\lambda$5007/H$\beta$ ratios  done by \citet{dopitaetal1984} in which they used fully-radiative plane-parallel shock models to estimate the [O\,{\sc iii}]$\lambda$5007/H$\beta$ ratio (among other elemental ratios) as a function of shock velocity. In \citet{dopitaetal1984} study their Figs.5 $\&$ 6, show the plots of this ratio as a function of shock velocity. Our [O\,{\sc iii}]$\lambda$5007/H$\beta$ ratio value of 4.5 obtained from our spectral analysis for SNR candidate ID 5 gives  us  an estimate of 100 < Vs < 150 km s$^{-1}$.
The interstellar extinction and the extinction of H$\alpha$ produced by such a dust screen distribution are also determined through the relations of \citet{relanoetal2006b, buatetal2002}:

\begin{align}
A_{(H\alpha)} = 5.25 \times  log[(H\alpha/H\beta)/(2.859\times t_e^{-0.07})]\\
E(B-V) = 0.44/2.45 \times A_{(H\alpha)}
\end{align}
where H$\alpha$/H$\beta$ stands for the observed H$\alpha$ and H$\beta$ emission line flux ratios and, the electron temperature, $t_{\rm e}$, is given in units of 10$^4$ K. For the calculation of Neutral Hydrogen column density, the relation from \citet{predehlandschmitt1995} was used

\begin{equation}
N(H\,${\sc i}$) = 5.4x10^{21} x E(B - V)
\end{equation}

The values calculated above for the spectrum of No.1 observations of SNR Candidate ID5 together with their parameters, the relative line fluxes, the electron density $N_{\rm e}$ (calculated with The Space Telescope Science Data Analysis System (STSDAS) task nebular.temden. program; this task is based on the program FIVEL \citep{derobetisetal1987, shawanddufour1995} for a five level atom) derived from the [S\,{\sc ii}]($\lambda$6716/$\lambda$6731) ratio for an assumed electron temperature 10$^4$ K \citep{osterbrockandferland2006} are shown in Table~\ref{tab:spectral_result}.

\begin{table*}
\centering
  \caption{Relative line intensity and the parameters obtained from the spectrum of the SNR candidate ID5. Fluxes are normalised  F(H$\alpha$) = 100. The signal-to-noise ratio are represent in parentheses beside their values and the errors of the emission line ratios and other parameters are calculated through standard error calculation.}
  \label{tab:spectral_result}
  \begin{tabular}{@{}cc|cc@{}}
  \hline
 Lines & Flux (S/N)& Parameters& Values\\
   &  & $\&$ & \\
 (\AA) & F(H$\alpha$=100) &Line Ratios & \\
 \hline
 H$\beta$ ($\lambda$ 4861)& 14.58 (4) & I (H$\alpha$) (erg cm$^{-2}$ s$^{-1}$) & (2.57 $\pm {0.01}$)$\times 10^{-14}$\\
 $[$O$\,${\sc iii}$]$ ($\lambda$ 4959) & 19.68 (7) & [S$\,${\sc ii}$]$$^a/$H$\alpha$ & 0.46 $\pm$ 0.01\\
 $[$O$\,${\sc iii}$]$ ($\lambda$ 5007) & 65.60 (18) & [S$\,${\sc ii}$]$$\lambda6716/\lambda$6731& 1.30 $\pm$ 0.03\\
 $[$N$\,${\sc ii}$]$ ($\lambda$ 6548) & 6.39 (6) & [O$\,${\sc iii}$]$ ($\lambda$ 5007)/H$\beta$ ($\lambda$ 4861) & 4.5 $\pm$ 0.07\\
 H$\alpha (\lambda$ 6563) & 100 (32) & $V_{\rm s}$ (km s$^{-1}$) & 100 - 150\\
 $[$N$\,${\sc ii}$]$ ($\lambda$ 6584) & 6.75 (7) & E(B-V)& 0.36 $\pm$ 0.01\\
 $[$S$\,${\sc ii}$]$ ($\lambda$ 6716) & 26.11 (4)& $A_{(H\alpha)}$ & 1.99 $\pm$ 0.03\\
$[$S$\,${\sc ii}$]$ ($\lambda$ 6731) & 20.10 (3) & N(H$\,${\sc i}$)$(cm$^{-2})$ & (0.19 $\pm {0.02})\times10^{22}$\\
  &  & N$_e$(cm$^{-3}$)& 121 $\pm$ 17\\
 \hline
\multicolumn{4}{l|}{$^a [S\,${\sc ii}$]$ is the combination of $\lambda$6716 and $\lambda$6731 flux values.}
\end{tabular}
\end{table*}

\subsection{X-ray Observation and Data Reduction}
We have analysed the $\sim$97 ks \textit{Chandra} X-ray Observatory Advanced CCD Imaging Spectrometer array (ACIS-S3) archival data of the galaxy NGC 1569 (ObsID: 782; PI: C. L. Martin) observed on 2000 April 11. The observation was telemetered in FAINT Data Mode. The data of this observation were reprocessed from Level 1 to Level 2, and analysed with \textit{Chandra} X-ray Center (CXC) \textsc{CIAO} 4.9\footnote{\url{ http://cxc.cfa.harvard.edu/ciao/}} package to remove pixel randomisation and to correct for CCD charge-transfer inefficiencies, and \textsc{CALDB} 4.7.6 was used for the most recent calibration products. Spectral analysis was performed with \textsc{xspec} \citep{arnaud1996} version 12.9.1\footnote{https://heasarc.gsfc.nasa.gov/docs/xanadu/xspec/index.html} with AtomDB v3.0.9\footnote{\url{ http://www.atomdb.org }} \citep{smith2001, foster2012}. In this study, we selected only those data in the energy range from 0.5 to 2 keV for the soft band emission, and 0.5 to 7 keV for the full band emission to have lower particle background and lower chip quantum efficiency contamination.

\subsection{X-ray Spectral Analysis}

%%%%% BACKGROUND ANALYSIS %%%%%%

SNR candidates in our sample have low surface-brightness as being extended sources in the X-ray band. Therefore, background modelling is required to obtain meaningful information from the source. As it is the matter of the background estimation, different responses and background contributions may be associated to different regions. As a first step, in order to estimate the astrophysical background emission, the background spectra were taken from a nearby region for each source. But this method resulted in a loss in the statistical quality of the source spectra. Alternatively, selecting from a nearby emission-free circular region (centered at $\rmn{RA}(\rm J2000)=04^{\rmn{h}} 30^{\rmn{m}} 39^{\rmn{s}}$, $\rmn{Dec.}~(\rm J2000)=64\degr 54\arcmin 37\arcsec$, with a radius of 74.46 arcsec), we defined a physically motivated background spectrum. We modelled it with various thermal and non-thermal spectral components (such as \textsc{apec} and \textsc{power-law}) and simulated it using \textsc{fakeit} command under the \textsc{xspec}, and then, subtracted this simulated background spectrum from the source spectrum. In this observation, the Cosmic X-ray Background (CXB) and the Galactic Foreground Emission (GFE) are considered as X-ray background components. Note that, as we did not select the X-ray background emission in line of sight of the host galaxy, the foreground emission is only associated with the Milky Way, not with the host galaxy, nor with their emission combination of the galaxies. As mentioned before, we did not selected an individual nearby background region for each point source spectrum since doing it this way it resulted in a loss in the statistical quality of the spectra. Otherwise, the total foreground emission would associated with both the Milky Way and the host galaxy. CXB was modelled with an absorbed power-law spectrum, with a photon index fixed at $\Gamma$ = 1.46 \citep{chen1997}. The normalisation is the free parameter because of the non-uniformity of this component \citep{maggi2016}. The GFE has generally small or no absorption (in our analysis it has yielded no absorption value) and modelled with \textsc{apec} model, which gave an \textit{kT$_{e}$} value of $\sim$0.24 keV \citep{ryu2009}. Aforementioned simulated background spectrum based on the background model, was normalised for the source region and subtracted from the spectra of the sources during the individual fitting process.\\

In the fitting process, the spectral models of the sources were coupled with the photoelectric absorption model (\textsc{phabs}) to account for the total column density by Galactic + intergalactic + host galaxy. This value, $N_{\rm H}$, was fixed at 0.19 $\times 10^{22}$ cm$^{-2}$, adapted from our calculation of Neutral Hydrogen column density mean value obtained from our optical observation (see Section 2.2, Eq. 3). \\

%%%%% SOURCE ANALYSIS %%%%%%

\begin{figure}
 \includegraphics[width=\columnwidth]{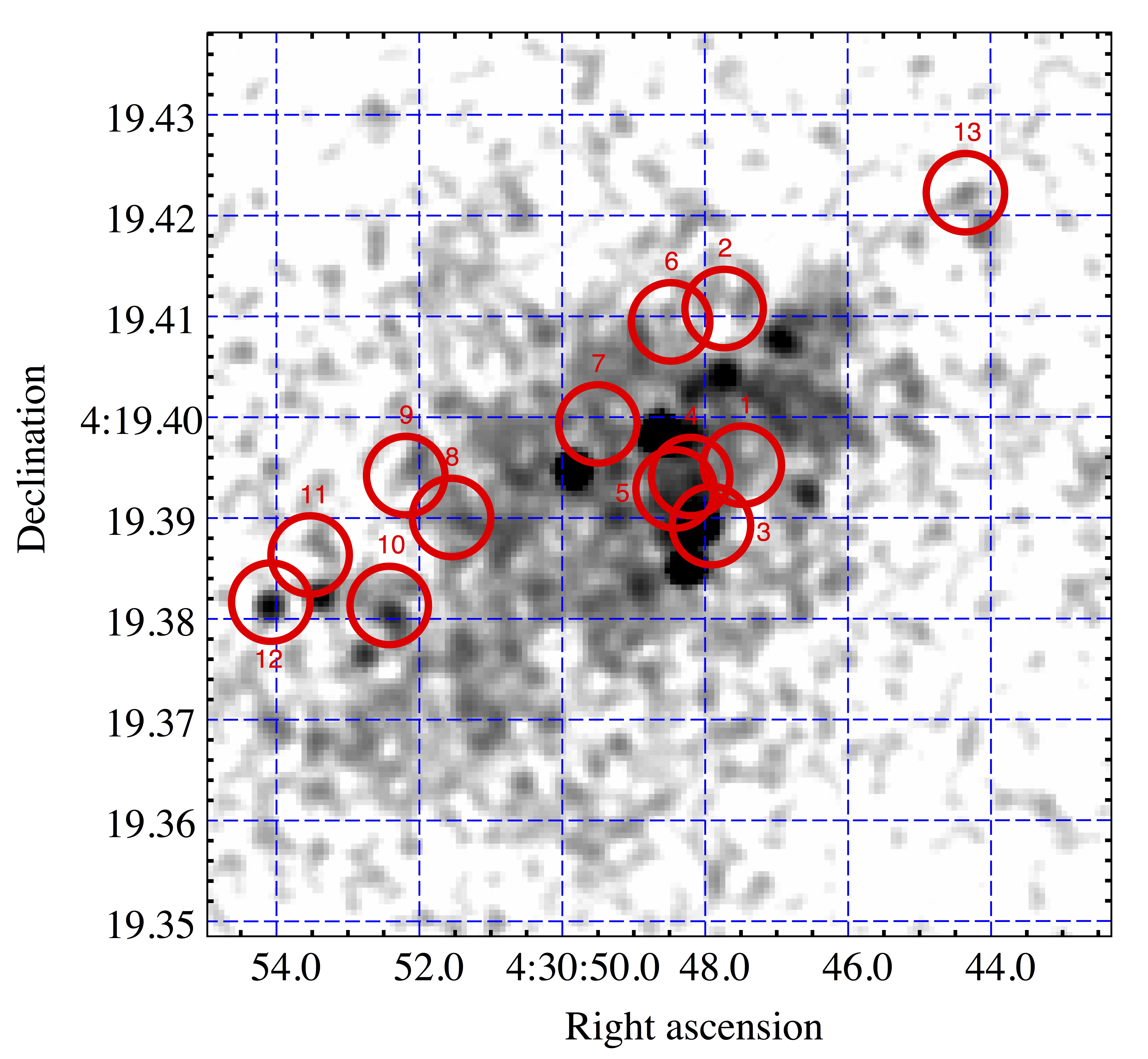}
 \caption{Exposure corrected \textit{Chandra} ACIS-S3 image of NGC 1569 in the 0.5$-$7.0 keV energy band. The numbered circles indicate the extracted regions for the spectral analysis.}
 \label{fig:ngc1569_regions}
\end{figure}

As a next step, we extracted the spectra of the circular region of each SNR candidates (3.5 arcsec in radius). We grouped the spectra with a minimum of 3 counts bin$^{-1}$. The more appropriate method in the case of the small number of counts than $\chi^{2}$ is modelling the spectral regions in order to allow the use of the Cash maximum likelihood statistic \citep{cash1979}. Also, all sources in this study are considered as point-like (3.5 arcsec, see Fig.~\ref{fig:ngc1569_regions}).

\begin{figure}
 \includegraphics[width=\columnwidth]{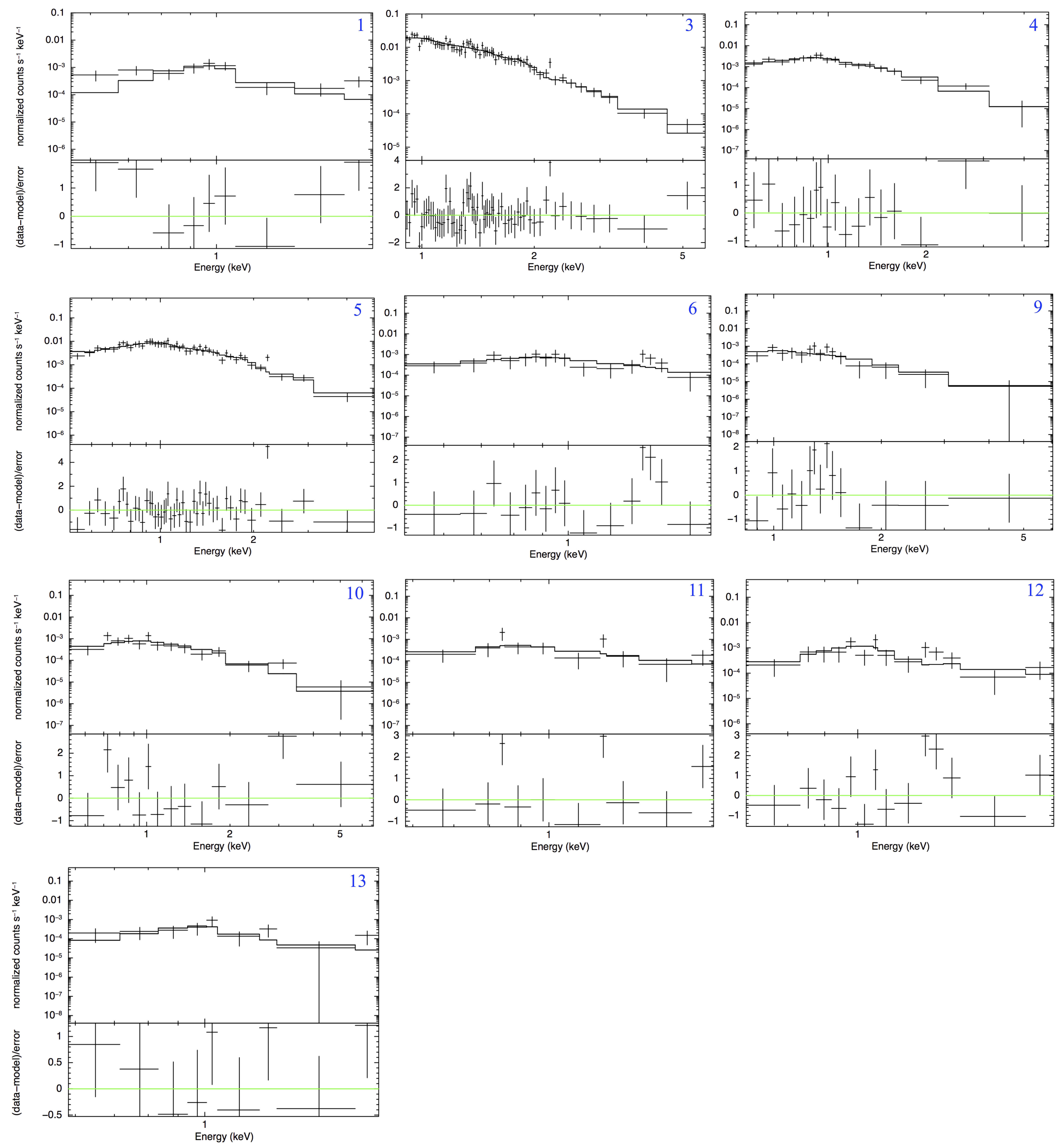}
 \caption{Background-subtracted ACIS spectra of NGC 1569 SNR candidates in the 0.5$-$2.0 and 0.5$-$7.0 keV energy bands, extracted from the point-like source regions, overlaid with the best-fitting model. The lower panels show the residuals. The spectral fitting results are summarised in Table~\ref{tab:Table_Xray}.}
 \label{fig:X-ray_spectra}
\end{figure}

In each spectral analysis, using the photoelectric absorption model of \textsc{phabs} \citep{balucinska1992, ott2005} to account for the total column density, and setting the elemental abundances to the values measured by \citet{anders1989}, we first tried to fit the spectra with a single-component Collisional Ionization Equilibrium (CIE) model \textsc{apec} or variable-abundance CIE model \textsc{vapec} to search for a visual inspection for evidence of line emission. According to the residuals and goodness-of-fit, we thawed the elemental abundances or switched to non-equilibrium ionisation (NEI) models \textsc{vnei} or \textsc{vpshock} in \textsc{xspec}. In line with the feature of each model, the free parameters were the normalisation, the absorption ($N_{\rm H}$), the electron temperature (\textit{kT$_{e}$}), and the ionisation timescale ($\tau$ = $n_{\rm e}$t, where $n_{\rm e}$, and \textit{t} represent the mean electron density, and the elapsed time after the shock heating of the plasma to a constant temperature \textit{kT$_{e}$}, respectively). The SNR candidates of ID4, ID5 and ID9 also have harder components to their spectra as showing emission above 5 keV, thus, we also tried to fit the spectra that show no line feature with \textsc{power-law} model in \textsc{xspec} to search for the non-thermal emission components, but their statistically unacceptable fits (C-stat/d.o.f $>$ 1.5) yielded to conclusion that the thermal model is more likely to be correct. We also tried to fit with two-temperature NEI or CIE models, and obtained no statistical improvement on the fits because of the insufficient count rates from the sources. Thus, signatures of two-temperature component of NEI or CIE cannot be attained in the terms of a statistically significant improvement, neither in the goodness-of-fit, nor in the residuals. On the other hand, most of the SNR candidates have soft spectrum in the 0.5$-$2 keV band, and almost show no significant emission features above 2 keV. Thus, we evaluated their spectra through their proportional contribution in the soft (0.5$-$2 keV) and full (0.5$-$7 keV) energy bands with the best-fitting values.\\
 As the ionisation ages $\tau$ in the samples were poorly constrained or high, using a NEI model did not result in a statistically significant improvement in the fits, thus, fitting with a CIE model was sufficient. Herewith, assuming \textsc{apec} model based on the quality of the fit is a more preferable way for a general approach, as most of the candidates are too faint to distinguish between CIE and NEI. Amongst the SNR candidates, three of them did not require or allow elemental abundances to be fitted, neither in the soft band, nor in the full band. This could be due to insufficient data, which retain the use of free abundances in the fits. Besides, we tried to fit the SNR candidates of ID2, ID7 and ID8 with both thermal and non-thermal models, but we could not get acceptable fits with unrealistic best-fitting parameters. Hence, these sources are denoted as unclassified. The spectral results and count rates can be seen in Table~\ref{tab:Table_Xray} and individual spectra are displayed in Fig.~\ref{fig:X-ray_spectra}. Using \textsc{xspec}, the count rates were determined from each individual source by extracting the background. We calculated the uncertainties at the 90 per cent confidence level with the \textsc{error} command of \textsc{xspec}.\\

We obtained the unabsorbed X-ray fluxes and luminosities in related bands from the best-fitting models (Table~\ref{tab:Table_flux}) with the \textsc{xspec} command \textsc{flux}. Using the distance to NGC 1569 \citep[3.4 Mpc;][]{kobulnickyandskillman1997, jacksonetal2011, lellietal2014}, we estimated the luminosity directly by our observed unabsorbed flux values, because of the relative certainty of the SNR candidates distances.\\

%%%%%%%%%%% Table_Xray %%%%%%%%%%%%%%%%%%%%%%%%%%%%%

\begin{table*}
\small
\begin{center}
  \caption{X-ray spectral results of NGC 1569 SNR candidates.}
   \label{tab:Table_Xray}
\begin{tabular}{llcccll}
\hline
SNR & X-ray Count rate & kT$_{e}$ & Abundances & norm & C-stat/d.o.f \\
Cand. ID & ($\times 10^{-4}$ count $^{-1}$) & (keV) &  ($10^{-2}$) & ($10^{-5}$ cm$^{-5}$) & \\
\hline

1$^{a}$         & 11.85 $\pm{\hspace{0.06cm}1.14}$   & 0.98 $\pm{\hspace{0.06cm}0.17}$   &	   (1)                                                                              & 0.11 $\pm{\hspace{0.06cm}0.03}$  & 14.53/16  \\    [0.13 cm]

3$^{a}$         & 163.50 $\pm{\hspace{0.06cm}4.12}$ & 1.09 $\pm{\hspace{0.06cm}0.12}$   & 3.47$\pm{\hspace{0.06cm}0.03}$                                   & 17.67$^{+3.89}_{-2.73}$                 & 110.06/104 \\ [0.13 cm]

4$^{b}$         & 26.62 $\pm{\hspace{0.06cm}1.68}$   & 0.95 $\pm{\hspace{0.06cm}0.15}$   & 5.03$^{+0.04}_{-0.08}$                                                   & 2.48$^{+0.71}_{-0.58}$   		  & 39.94/48	\\ [0.13 cm]

5$^{b}$        & 89.12 $\pm{\hspace{0.06cm}3.05}$    & 1.07$^{+0.13}_{-0.08}$                    & 3.24$^{+0.03}_{-0.02}$                                                   & 9.39$^{+1.37}_{-1.44}$                    & 105.42/103 \\  [0.13 cm]

6$^{a}$        & 11.33 $\pm{\hspace{0.06cm}1.12}$     & 0.91$^{+0.99}_{-0.26}$                    & 5.23 $\pm{\hspace{0.06cm}0.27}$                                 & 0.66$^{+0.45}_{-0.31}$                   & 10.32/13	 \\  [0.13 cm]

9$^{b}$        & 4.43 $\pm{\hspace{0.06cm}0.72}$       & 1.36$^{+0.82}_{-0.68}$                    & 3.48 $\pm{\hspace{0.06cm}0.13}$                                 & 0.58$^{+0.72}_{-0.32}$ 		   & 10.72/12	\\  [0.13 cm]

10$^{a}$      & 12.36 $\pm{\hspace{0.06cm}1.16}$     & 1.18$^{+0.68}_{-0.33}$                      & 2.22 $\pm{\hspace{0.06cm}0.28}$                               & 0.91$^{+0.41}_{-0.26}$     		   & 23.63/24	\\  [0.13 cm]

11$^{a}$      & 3.94 $\pm{\hspace{0.06cm}0.69}$       & 0.84$^{+0.44}_{-0.28}$                      & 4.69 $\pm{\hspace{0.06cm}0.19}$                                & 0.44$^{+0.53}_{-0.32}$	            & 7.19/7	\\  [0.13 cm]

12$^{a}$      & 10.92 $\pm{\hspace{0.06cm}1.10}$     & 1.18 $\pm{\hspace{0.06cm}0.18}$     & 		    (1)                                                                    & 0.12 $\pm{\hspace{0.06cm}0.05}$  & 13.75/12\\ [0.13 cm]

13$^{a}$      & 9.06 $\pm{\hspace{0.06cm}1.01}$       & 1.04$^{+0.48}_{-0.33}$                      &   		     (1)                                                           & 0.05 $\pm{\hspace{0.06cm}0.03}$   & 5.46/7 \\

\hline
\end{tabular}
\begin{flushleft}
\textit{\rm Note: The normalisation, norm=$10^{-14}$$\int n_{\rm e} n_{\rm H} dV$/($4\pi d^{2}$) (cm$^{-5}$), where $n_{\rm e}$, $n_{\rm H}$, $V$ and $d$ are the electron and hydrogen densities (cm$^{-3}$), emitting volume (cm$^{3}$) and distance to the source (cm), respectively. $^{a}$: 0.5$-$2 keV, $^{b}$: 0.5$-$7 keV energy band. (1) indicates that the elemental abundance was fixed at solar.}
\end{flushleft}
\end{center}
\end{table*}

%%%%%%%%%%% Table_flux %%%%%%%%%%%%%%%%%%%%%%%%%%%%%
\begin{table*}
\small
\begin{center}
  \caption{Unabsorbed X-ray fluxes and luminosity values of NGC 1569 SNR candidates, considering the distance of 3.4 Mpc \citep{lellietal2014, jacksonetal2011, kobulnickyandskillman1997}.}
  \label{tab:Table_flux}
\begin{tabular}{lll}
\hline \\[-2.0ex]
SNR         & F$_{X}$                                                          & L$_{X}$ \\
Cand. ID  & ($10^{-15}$ erg cm$^{-2}$ s$^{-1}$)     & ($10^{36}$ erg s$^{-1})$    \\ [0.14 cm]
\hline \\[-2.0ex]

1$^{a}$    & 1.55$^{+0.50}_{-0.30}$                                   & 2.14$^{+0.69}_{-0.41}$  \\ [0.14 cm]
3$^{a}$    & 58.34$\pm{6.31}$                                           &  80.51$\pm{8.71}$\\ [0.14 cm]
4$^{b}$    & 7.80$^{+0.94}_{-1.48}$                                   &  10.76$^{+1.30}_{-2.04}$\\ [0.14 cm]
5$^{b}$    & 30.88$^{+1.70}_{-2.88}$                                 &  42.61$^{+2.35}_{-3.67}$\\ [0.14 cm]
6$^{a}$    & 1.62$^{+0.35}_{-0.61}$                                   &  2.24$^{+0.48}_{-0.84}$\\ [0.14 cm]
9$^{b}$    & 2.70$^{+0.89}_{-1.67}$                                   &  3.73$^{+1.23}_{-2.30}$\\ [0.14 cm]
10$^{a}$  & 2.15$^{+0.80}_{-1.34}$                                   &  2.97$^{+1.10}_{-1.85}$\\ [0.14 cm]
11$^{a}$  & 1.06$^{+0.34}_{-0.70}$                                   &  1.46$^{+0.47}_{-0.97}$\\ [0.14 cm]
12$^{a}$  & 1.06$^{+0.41}_{-0.81}$                                  &   1.46$^{+0.56}_{-1.12}$\\ [0.14 cm]
13$^{a}$  & 0.58$\pm{0.25}$                                             &  0.80$\pm{0.34}$\\ [0.14 cm]

\hline
\end{tabular}
\begin{flushleft}
{\hspace{34pt}$^{a}$: 0.5$-$2 keV, $^{b}$: 0.5$-$7 keV energy band.}
\end{flushleft}
\end{center}
\end{table*}

\section{DISCUSSION}

\subsection{Optical imaging and spectral results}

In this work, the imaging and spectroscopic observations of NGC 1569 are obtained in optical band by using RTT150 telescope.

This research is based on the criteria of [S\,{\sc ii}]/H$\alpha$ $\geq$ 0.4, as suggested by \citet{blairandlong1997, mathewsonandclarke1973, dopita1997}, which is used for 13 SNR candidates in NGC 1569. These 13 SNR candidates with appropriate [S\,{\sc ii}]/H$\alpha$ ratios and their errors are presented in Table~\ref{tab:optic_result}.

\citet{chomiukandwilcots2009} reported their work in radio band and categorized SNR candidates in NGC 1569 by using spectral indices and H$\alpha$ images. 23 SNR candidates were reported in their work for NGC 1569 at its central region. In this work, out of 13 SNR candidates reported here, 8 of them (i.e. with ID1, 2, 3, 4, 5, 6, 7, 9, 12, 13 see Table~\ref{tab:optic_result}) were found to be consistent with \citet{chomiukandwilcots2009} results.

The SNR candidate with ID7 shown in our Table~\ref{tab:optic_result}, is also reported by \citet{greveetal2002}. Throughout NGC 1569 they reported 4 to 5 SNe and/or SNRs, located within an area of approximately 300 pc diameter around their super star clusters A, B, C where active star formation occurred until recently, and might still be going on.

In order to check the photometry calibration, H$\alpha$ fluxes of our 13 SNR candidates were compared as shown in Table~\ref{tab:optic_result} and marked if they were earlier detected ones, as mentioned before. H$\alpha$ flux values are found to be around (2.2--32) $\times$ 10$^{-15}$ erg cm$^{-2}$ s$^{-1}$.

Only the spectrum of SNR candidate ID5 observations was available for analysis in this work; since the observations were done only for ID5, because of limited seeing conditions during observations, as mentioned in the previous section. This SNR candidate with ID5 here, corresponds to the coordinates given by \citet{chomiukandwilcots2009} at ($\rmn{RA}(\rm J2000)=04^{\rmn{h}} 30^{\rmn{m}} 48^{\rmn{s}}$.42, $\rmn{Dec.}~(\rm J2000)=64\degr 50\arcmin 53\arcsec$.60) which is in a good agreement with ours. ID5$'$s spectrum was obtained with the low-resolution RTT150 TFOSC. The observed emission line spectrum of the SNR candidates contains typical permitted lines of H$\alpha$, H$\beta$ and forbidden lines regions where electron density is low, this ratio should be approximately 1.5, as was pointed out by \citet{franketal2002}. In this study, this value is found to be 1.3 which corresponds to the electron density value, $N_{\rm e}$, of about 121 $\pm$ 17 cm$^{-3}$ (Table~\ref{tab:spectral_result}).

Following the work done by \citet{dopitaetal1984} and their Fig5. and Fig.6, the shock wave velocity, $V_{\rm s}$, is estimated from their [O\,{\sc iii}]$\lambda$5007/H$\beta$ line ratio diagnostics. Our optical value of ([O\,{\sc iii}]$\lambda$5007/H$\beta$) = 4.5 is in a very good agreement with their estimate of 100 < Vs < 150 km s$^{-1}$. The hydrogen column density of the SNR candidate ID5 was found to be (0.19 $\pm$ 0.02) $\times10^{22}$ cm$^{-2}$.

\subsection{X-ray properties of the optically selected SNR candidates}

We examined the X-ray properties of the optically detected possible SNR candidates in this work, using the thermal plasma models to search for evidence of the thermal emission, since most SNR candidates are extended sources at X-ray wavelengths, and their spectra are dominated by emission from a hot plasma. Non-thermal X-ray emission is not required to describe the observed X-ray spectrum for the 13 possible SNR candidates. Out of the 13 sources, only 10 of them have been suggested to be possible SNR candidates based on their thermal X-ray spectra, and the remaining three are denoted as unclassified because of their unacceptable fit parameters. We detected low elemental abundances in seven sources listed in Table~\ref{tab:Table_Xray}. Based on their spectra, we measure the unabsorbed fluxes with a range of (0.58 $\lesssim$ \textit{F$_{X}$} $\lesssim$ 58.43) $\times$ $10^{-15}$ erg cm$^{-2}$ s$^{-1}$, and the unabsorbed luminosities with a range of (0.80 $\lesssim$ \textit{L$_{X}$} $\lesssim$ 80.51) $\times$ $10^{36}$ erg s$^{-1}$ for 10 X-ray sources. As seen in Table~\ref{tab:Table_flux}, the SNR candidates ID3 and ID5 have very high luminosity values compared with the others. The faintest source is SNR candidate ID13, which is located at the outer rim of NGC 1569. Since most of the candidates are too faint to distinguish between CIE and NEI, even the brightest sources SNR candidates ID3 and ID5, assuming CIE model \textsc{apec} is a more preferable way for a general approach to understand the conditions of the plasma. The spectral analysis revealed that spectra of the SNR candidates are best modelled with the CIE plasma model with a temperature range of 0.84 keV $\lesssim$ \textit{kT$_{e}$} $\lesssim$ 1.36 keV, since the soft spectrum with \textit{kT$_{e}$} $\leq$ 2 keV, which is a characteristic of the thermal SNRs, and this situation limits the possibility that they are background sources (e.g., active galactic nuclei) as indicated by \citet{leonidaki2010}. Similarly, the best-fitting spectral parameters with the large errors and having small number of counts point out the fact that the considered sources might be an SNR \citep{leonidaki2010}.

\citet{sánchez-crucesetal2015} have reported 54 possible X-ray source coordinates as shown in their Table 9. They listed 54 of them with their possible description of their characteristic types that they considered as  active galactic nuclei, X-ray sources, X-ray binaries, stars and two SNR candidates. On the other hand, when we have compared our coordinates originated from our optical and X-ray data analysis results for the possible SNR candidates reported here, in our work and also the radio observations reported by \citet{chomiukandwilcots2009} any of their two possible SNR candidates did not match.

\section{CONCLUSIONS}

In this work, our aim is to search for the possible SNR candidates in optical and X-ray wavelengths in NGC 1569. Therefore, our conclusions are as follows:

\begin{enumerate}

\item An optical imaging survey of SNR candidates in NGC 1569 is presented with the help of RTT150 observations for the first time using the [S\,{\sc ii}]/H$\alpha$ technique on the optical narrowband images. This search is realised for possible SNR candidates in the nearby galaxy NGC 1569 by using the criteria of [S\,{\sc ii}]/H$\alpha$ ratio is $\geq$0.4 as mentioned earlier. Based on optical narrowband [S\,{\sc ii}] and H$\alpha$ imaging, we identified a total 13 SNR candidates in this galaxy.

\item The optical spectral observations were also performed for SNR candidate-ID5 which is obtained with the  RTT150 TFOSC. H$\alpha$, H$\beta$, [O\,{\sc iii}]$\lambda$4959, $\lambda$5007, [N\,{\sc ii}]$\lambda$6548, $\lambda$6584 and [S\,{\sc ii}]$\lambda$6716, $\lambda$6731 emission lines are shown in the spectrum in Fig.~\ref{fig:optical_spectrum}. The observed relative line fluxes and the parameters obtained from the spectrum are [S\,{\sc ii}]/H$\alpha$ = 0.46 $\pm$ 0.01, [S\,{\sc ii}]($\lambda$6716/$\lambda$6731) = 1.30  $\pm$ 0.03, shock wave velocity of 100 < Vs < 150 km s$^{-1}$ (with [O\,{\sc iii}]$\lambda$5007/H$\beta$), optical extinction E(B-V) = 0.36  $\pm$ 0.01, $A_{(H\alpha)}$ = 1.99  $\pm$ 0.03, absorbing column density N(H\,{\sc i}) = (0.19  $\pm$ 0.02) $\times10^{22}$ cm$^{-2}$ and electron density $N_{\rm e}$ = 121  $\pm$ 17 cm$^{-3}$ (with [S\,{\sc ii}]($\lambda$6716/$\lambda$6731)).

\item We investigated the X-ray properties of the optically detected possible SNR candidates using \textit{Chandra} archival data of NGC 1569. Out of 13 sources, only 10 of them have yielded best-fit parameters with good statistics. As seen in Table~\ref{tab:Table_Xray}, we measured the 0.5$-$2 keV and 0.5$-$7 keV band unabsorbed fluxes with a range of (0.58 $\lesssim$ \textit{F$_{X}$} $\lesssim$ 58.43) $\times$ $10^{-15}$ erg cm$^{-2}$ s$^{-1}$ for 10 X-ray sources. Our spectral analysis revealed that spectra of the SNR candidates are best modelled with the CIE plasma model with a temperature range of 0.84 keV < \textit{kT$_{e}$} < 1.36 keV.

\end{enumerate}

\section*{ACKNOWLEDGMENTS}

We are grateful to the referee for the insight and detailed comments that helped improve the manuscript. We thank TUBITAK for partial support in using RTT150 (Russian-Turkish 1.5-m telescope in Antalya) with project number 15BRTT150-842. This work is financially supported by Bogazici University, BAP through project number 8563. We thank Dr. Aytap Sezer for many useful discussions and critical reading of the manuscript and Dr. G. Branduardi-Raymont for her valuable comments.

\onecolumn

\twocolumn

\onecolumn
\twocolumn

\onecolumn

% Don't change these lines
\bsp	% typesetting comment
\label{lastpage}
\end{document}